\documentclass[aps,prl,twocolumn,groupedaddress,showpacs]{revtex4-1}
 \usepackage[pdftex]{graphicx}
\renewcommand{\figurename}{\textbf{Figure}}
\renewcommand{\thefigure}{\textbf{\arabic{figure}}}
\usepackage{ulem}

\begin{document}
\title{Spin Chains and Electron Transfer at Stepped Silicon Surfaces}
\author{J. Aulbach$^1$, S. C. Erwin$^2$, R. Claessen$^1$ and J. Sch\"afer$^1$}
\affiliation{$^1$Physikalisches Institut and R\"ontgen Center for Complex Material Systems (RCCM), Universit\"at W\"urzburg, D-97074 W\"urzburg, Germany\\
$^2$Center for Computational Materials Science, Naval Research Laboratory, Washington, DC 20375, USA}

\begin{abstract} High-index surfaces of silicon with adsorbed gold can
reconstruct to form highly ordered linear step arrays. These steps
take the form of a narrow strip of graphitic silicon. In some
cases---specifically, for Si(553)-Au and Si(557)-Au---a large fraction
of the silicon atoms at the exposed edge of this strip are known to be
spin-polarized and charge-ordered along the edge. The periodicity of
this charge ordering is always commensurate with the structural
periodicity along the step edge and hence leads to highly ordered
arrays of local magnetic moments that can be regarded as ``spin
chains.''  Here, we demonstrate theoretically as well as
experimentally that the closely related Si(775)-Au surface
has---despite its very similar overall structure---zero spin
polarization at its step edge.  Using a combination of
density-functional theory and scanning tunneling microscopy, we
propose an electron-counting model that accounts for these
differences.  The model also predicts that unintentional defects and
intentional dopants can create local spin moments at Si($hhk$)-Au step
edges. We analyze in detail one of these predictions and verify it
experimentally.  This finding opens the door to using techniques of
surface chemistry and atom manipulation to create and control silicon
spin chains.
\end{abstract}

 \maketitle

Understanding how magnetism arises in materials without $d$ electrons
poses a scientific challenge for both theory and experiment. Such an
understanding also offers a tantalizing technological goal, namely
the integration of semiconductor properties and nonvolatile magnetism
in a single material system. One path toward this goal, based on
dilute magnetic semiconductors, has developed over the past two
decades. Another approach is much more recent, and exploits instead
specific features of the underlying crystal structure of the
semiconductor. For example, in both carbon and silicon the electronic
properties of orbitals not in covalent bonds---that is, partially
filled ``dangling bond'' orbitals---can give rise under certain
circumstances to magnetic states. These states, and the particular
conditions that lead to magnetism, are just beginning to be
investigated systematically.

Carbon graphene nanoribbons provide a prominent example of magnetic
states at the unpassivated edge of an otherwise non-magnetic material
\cite{NatureGraphen, Li2009, thiago2008, chih2004}.  For silicon the
situation is more complicated. Extended graphitic silicon does not
exist naturally. But nanoscale graphitic silicon ``ribbons'' do exist
and indeed form by self-assembly on stepped silicon substrates. The
unpassivated edges of these graphitic steps are, in some cases,
spin-polarized and perhaps even magnetically ordered at very low
temperature. The best studied such example is Si(553)-Au, a stepped
surface created when a submonolayer amount of gold is incorporated
into the first atomic layer of a silicon substrate
\cite{Crain2004}. At low temperatures the step edges of Si(553)-Au
develop a tripled periodicity which is now understood to arise from
the complete spin polarization of every third dangling bond along the
step edge
\cite{Ahn2005,Snijders2006,Krawiec,Erwin2010,Voegeli,Snijders2012,Aulbach}.

The Si(553)-Au system is one member of a family of Au-stabilized Si
surfaces miscut from (111) toward or away from the (001)
direction. This family is hence collectively denoted as
Si($hhk$)-Au. Each member of the family (four are known and others may
well exist) is built from the small set of recurring structural motifs
depicted in Figure\ 1. Each has a basic terrace-plus-step structure. The
step edge is always a single-honeycomb graphitic strip of silicon. The
terrace width varies according to the Miller index ($hhk$). A chain of
gold atoms, either one or two atoms wide, is incorporated on each
terrace. For sufficiently wide terraces a row of silicon adatoms is also
present. Notwithstanding these strong familial similarities, the
individual members of the family differ in one important respect: the
step edges of {\it some} members form spin chains---and
hence may order magnetically at low temperature---while others do not.

\begin{figure}
\includegraphics[width=.99\linewidth]{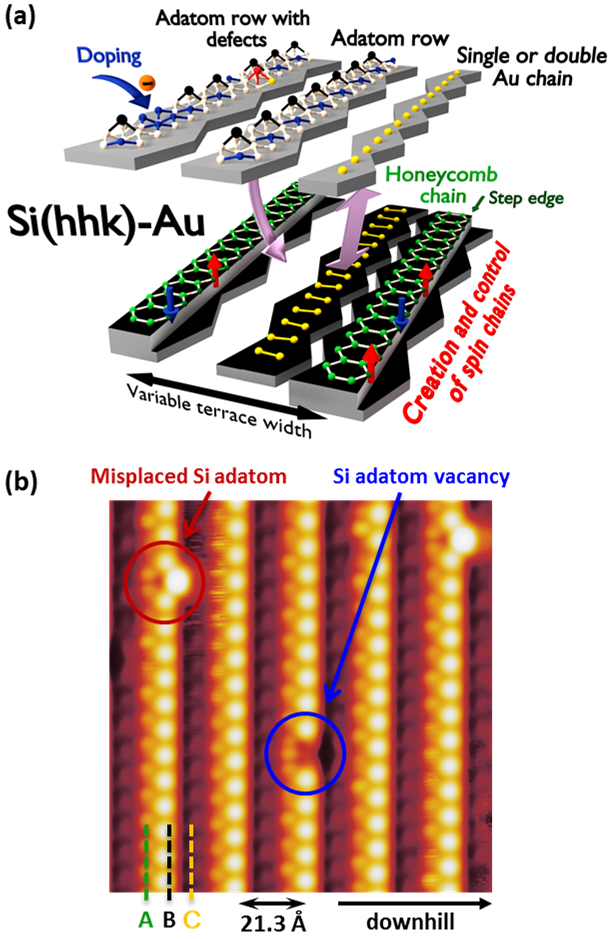}
\caption{(a) Overview of the important structural motifs that control
the formation of spin chains at the silicon honeycomb step edges (green hexagons) of
Si($hhk$)-Au surfaces. Different crystallographic orientations ($hhk$)
alter the terrace width, which leads to variations in the width of the
Au chain (single or double) and to the presence or absence of
additional rows of Si adatoms, which in turn open the door to adatom
defects. The Au chains, Si adatoms, and defects collectively
determine the electronic configuration of the silicon dangling bonds
at the step edge, and hence governs the formation or suppression of 
local spin moments.  (b) Topographic STM image of empty states on
Si(775)-Au at 77 K, 150 pA, $+$1.0 V. Three distinct rows
(A, B, C) are labeled. Two common native defects are labeled and discussed in the text.}
\label{fig:Topo}
\end{figure}

In this Letter, we use scanning tunneling microscopy (STM) and
density-functional theory (DFT) to develop a physically transparent
picture explaining the formation of local spin moments in the family
of Si($hhk$)-Au surfaces.  Specifically, we explain under which
conditions spin chains are formed at the step edges of particular
silicon ($hhk$) orientations. We use the term {\it spin chain} to mean a linear array of local
magnetic moments with a high degree of positional order, well-defined
periodicity, and non-negligible interaction between the spins. The
question of whether such a spin chain becomes magnetically ordered is
a separate one, which we defer to later investigation. Our detailed analysis allows us to
suggest chemical pathways, such as the use of dopants, to control
formation or suppression of local spin moments, which are a
prerequisite for spin chains to form. As an example we demonstrate
that a common native defect---a surface vacancy---creates a single
spin at the step edge.

We develop our picture in several stages. (1) We use STM and DFT to
propose the first detailed structural model for Si(775)-Au, a gold
chain system closely related to the previously studied systems
Si(553)-Au and Si(557)-Au. (2) We demonstrate that although the atomic
structure of Si(775)-Au is very similar to those earlier systems, the
atoms comprising its step edge are {\it not} spin-polarized, and hence
that spin chains are not formed on Si(775)-Au. (3) We provide experimental
evidence supporting a theoretical prediction, made previously, that every second
atom along the step edges of Si(557)-Au is spin-polarized
\cite{Erwin2010}, and hence that spin chains do form on Si(557)-Au.
(4) We propose an electron counting model that explains these findings
and use this model to predict that atoms at the Si(775)-Au step edge
become spin-polarized when holes are added to the system. (5) We
identify a specific, commonly observed native defect on Si(775)-Au
which creates
 holes in the system and, indeed, renders the adjacent step edge atom 
 spin-polarized. Our findings open the door to using surface
chemistry---that is, depositing specific atoms or molecules on the
surface---in order to create, enhance, or suppress spin chains in
Si($hhk$)-Au systems.

{\bf Background and Preliminaries.} Before turning to the family of
complex high-index Si($hhk$)-Au surfaces sketched in Figure\ 1a, we
first set the stage by introducing a concept which is key to
understanding silicon spin chains. We consider the electronic
structure of the unusual graphitic honeycomb ribbon which forms the
step edge of all known Si($hhk$)-Au surfaces. This ribbon was first
addressed theoretically in Ref.\ \onlinecite{Erwin1998}, where it was
introduced as the ``honeycomb'' of the honeycomb chain-channel
3$\times$1 reconstruction of Si(111)-$M$, where $M$ is a metal
adsorbate. When this metal is an alkali element the system is a normal
band insulator. This is because, in the language of band theory, there
are exactly enough electrons (four, including one from the alkali
atom) to fully occupy the two band states (denoted $S_2^+$ and
$S_2^-$) formed from the two $sp^3$ orbitals belonging to the two
outer atoms of the honeycomb. In the language of dangling bond
orbitals, this is equivalent to saying that both of the outer atoms
have doubly occupied ``lone pairs'' of electrons. The basic electronic
configuration of the honeycomb in Si(111)-$M$ is hence insulating and
non-spin-polarized.

This same basic electron configuration also describes the ``ideal''
step edge of Si($hhk$)-Au surfaces---that is, before any electron
transfer to or from the terrace occurs. Indeed, our DFT calculations show that the
Au atom plays the role of the alkali atom by donating
one electron to the honeycomb step-edge
states. Hence we arrive at the
conclusion that the basic electronic configuration of the Si($hhk$)-Au
step edge is likewise insulating and non-spin-polarized, at least
before electron transfer is taken into account. The research presented
below addresses the various ways, both intrinsic and extrinsic, that
electrons can be transferred out of these step-edge states. This
transfer is prerequisite for the formation of silicon spin chains.

\begin{figure*}
\includegraphics[width=0.9\linewidth]{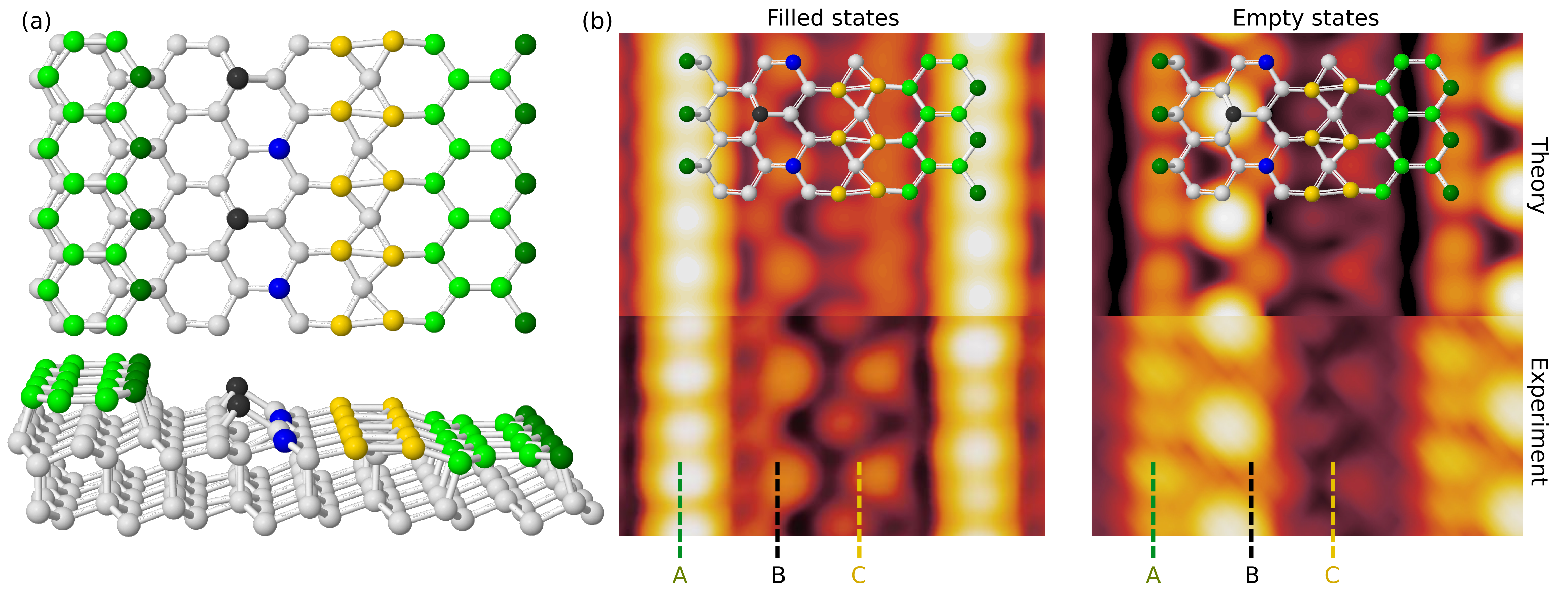} \caption{(a)
Proposed structure of Si(775)-Au. Yellow atoms are Au, all others are
Si. Each terrace contains a Au double row and a graphitic Si honeycomb
chain (green) at the step edge. The ladder structure of the Au row has
doubled (2$a_{0}$) periodicity due to an alternating twist of the
ladder rungs. A row of Si adatoms (black) with 2$a_{0}$ periodicity
passivates three surface dangling bonds per 1$\times$2 cell, leaving
one unpassivated Si restatom (blue) per 1$\times$2 cell. All of these
effects give rise to discernible features in the STM topography. (b)
Comparison of experimental (lower) and simulated (upper) STM images
for filled and empty states (experimental bias $-1.0$ and $+1.0$ eV,
theoretical bias $-0.8$ and $+0.8$ eV). Topographic features are
discussed in the text. Rows A, B, and C are marked as in Figure\ 1b
and also denote the locations at which the spectra in Figure\ 3a were
acquired.}  \label{fig:Model} 
\end{figure*}

{\bf Structural Model for Si(775)-Au.} The formation of local spin moments
at the steps of Si($hhk$)-Au systems depends on the electronic
configuration of the unpassivated
atoms at the step edge. Thus we begin by exploring an issue central to
step-edge magnetism: how many electrons are available to occupy these
orbitals? The answer is found on the terrace and, in particular, in
the structural motifs appearing there. Wide terraces have more motifs,
narrow terraces have fewer. The width of the terrace is determined by
the Miller index of the surface. This index $(hhk)$ can always be
written in the form $(h,h,h\pm2n)$ where $n$ is an integer. The unit
cell period $L$ is then given, in units of the silicon lattice
constant, by $L^2=S^2+T^2$ where $S=|h-k|/2\sqrt{3}$ is the step
height and $T=(2h/3+k/3)\sqrt{3/8}$ is the terrace width.  It is hence
evident that for surface orientations close to (111), larger values of
$h$ and $k$ correspond to wider terraces.

The terraces on Si(775)-Au are thus relatively wide, $T = 21.1$ \AA,
half again as wide as on Si(553)-Au. This additional space is
equivalent to two additional silicon unit cells and hence opens the
door to structural motifs not found on Si(553)-Au. We turn first to
the task of identifying these motifs using STM topographic imagery and
DFT calculations. The resulting structural model for Si(775)-Au will
then allow us to address in detail the electronic configuration---and
thus the issue of magnetism---at the step edge.

Figure 1b shows a constant-current STM image of
the Si(775)-Au surface. This image reveals three rows of features (A, B, C)
 separated by 21.3 \AA, as well as occasional random defects which we 
identify as Si-adatom vacancies and misplaced Si adatoms.  
The three rows each show a doubled periodicity with 
respect to the silicon surface lattice
constant $a_{0}=3.84$ \AA. 
These 1$\times$2 patterns do not change, within any row, in the
temperature range studied (between 5 K and 300 K; see Supporting Information
Figure\ S1). Hence there is no transition to higher order periodicity,
in contrast to the case of Si(553)-Au \cite{Ahn2005,Snijders2006,Erwin2013}. These images are also
consistent with previous STM investigations
\cite{Crain2004, Pedri}, although improved resolution now provides
tighter constraints on a detailed atomic model.

Figure 2a presents our proposed structural model of
Si(775)-Au. All of the motifs are taken from Figure\ 1a. The step edge
is a graphitic Si honeycomb chain, just as for all other members of
this family. On the terrace, a double chain of Au atoms occupies the
topmost Si layer and repairs the surface stacking fault created by the
graphitic chain. Hence the remainder of the terrace has a standard
Si(111) crystal structure. The bare Si(111) surface would be
energetically costly because of its high density of dangling bonds and
thus, on extended regions, reconstructs in the well-known
dimer-adatom-stacking-fault pattern. Such a reconstruction is not
possible on Si(775)-Au and hence this (111)-like region simply
incorporates Si adatoms to passivate most of its surface dangling
bonds. This creates a staggered double row of alternating adatoms and
unpassivated ``restatoms'' with 1$\times$2 periodicity.

Figure 2b compares theoretically simulated STM images
for this structural model to high-resolution experimental data. The
agreement is excellent in both the filled- and empty-state images and
allows us to identify the atomic origin of all experimentally observed
features. The bright rows visible in the occupied states, labeled A in
Figure\ 1, arise from the unpassivated atoms at the edge of the
graphitic step. The zigzag row just to the right of this step, labeled
B in Figure\ 1, arises from the staggered row of adatoms and
restatoms. The experimental bias dependence of both rows is striking
and in excellent agreement with theory: row A is bright in filled
states and weak in empty states, while in row B different parts of the
zigzag chain are highlighted by reversing the bias---the restatoms are
brightest in filled states, while the adatoms dominate the empty
states.

Dimerization of the Au double row gives
rise in STM to a double row of staggered spots. Specifically, a
1$\times$2 periodicity within each of the two rows is created because
the rungs of the Au ladder twist with alternating sign along the row
\footnote{The magnitude of this dimerization is sensitive to the
choice of DFT exchange-correlation functional and to the silicon
lattice constant. We find using PBE that the dimerization parameter $d
=(a_1-a_0)/a_0$ has the value 0.08.}.  Row C arises from the
right-hand leg of the ladder formed by the Au chain. An additional row
of weak spots, between rows B and C, arises from the left-hand leg in
similar fashion.

In summary, the detailed structure, spacing, periodicity, and bias
dependence of all the features observed experimentally by STM on
Si(775)-Au are accurately explained by our proposed structural model.

{\bf Absence of Spin Polarization in Si(775)-Au.} Having established a
plausible complete atomic model for Si(775)-Au, we turn now to its
electronic structure. Our main theoretical finding is that the step
edge is completely non-spin-polarized within DFT.  This is perhaps
surprising, especially in light of the many structural similarities to
Si(553)-Au, for which every third step-edge atom is completely
spin-polarized. The reason for this difference is that there are
enough electrons available to completely fill all the dangling-bond
orbitals at the Si(775)-Au step edge, thus forming doubly occupied
lone pairs with zero spin polarization. We will return to this
point below.

Experimentally, we used scanning tunneling spectroscopy (STS) with a lock-in technique to obtain a
direct measurement of the local electronic states at the step
edge. Figure 3a shows scanning tunneling spectroscopy
(STS) $dI/dV$ spectra taken at the locations marked 
in Figure\ 2b.  The spectrum taken at the step edge is
strongly peaked at $-$0.8 eV and has no significant weight above the
Fermi level. This is the signature of a fully occupied state and is
hence consistent with the theoretical result of a non-spin-polarized
lone pair at the Si(775)-Au step edge.

\begin{figure}
\includegraphics[width=.99\linewidth]{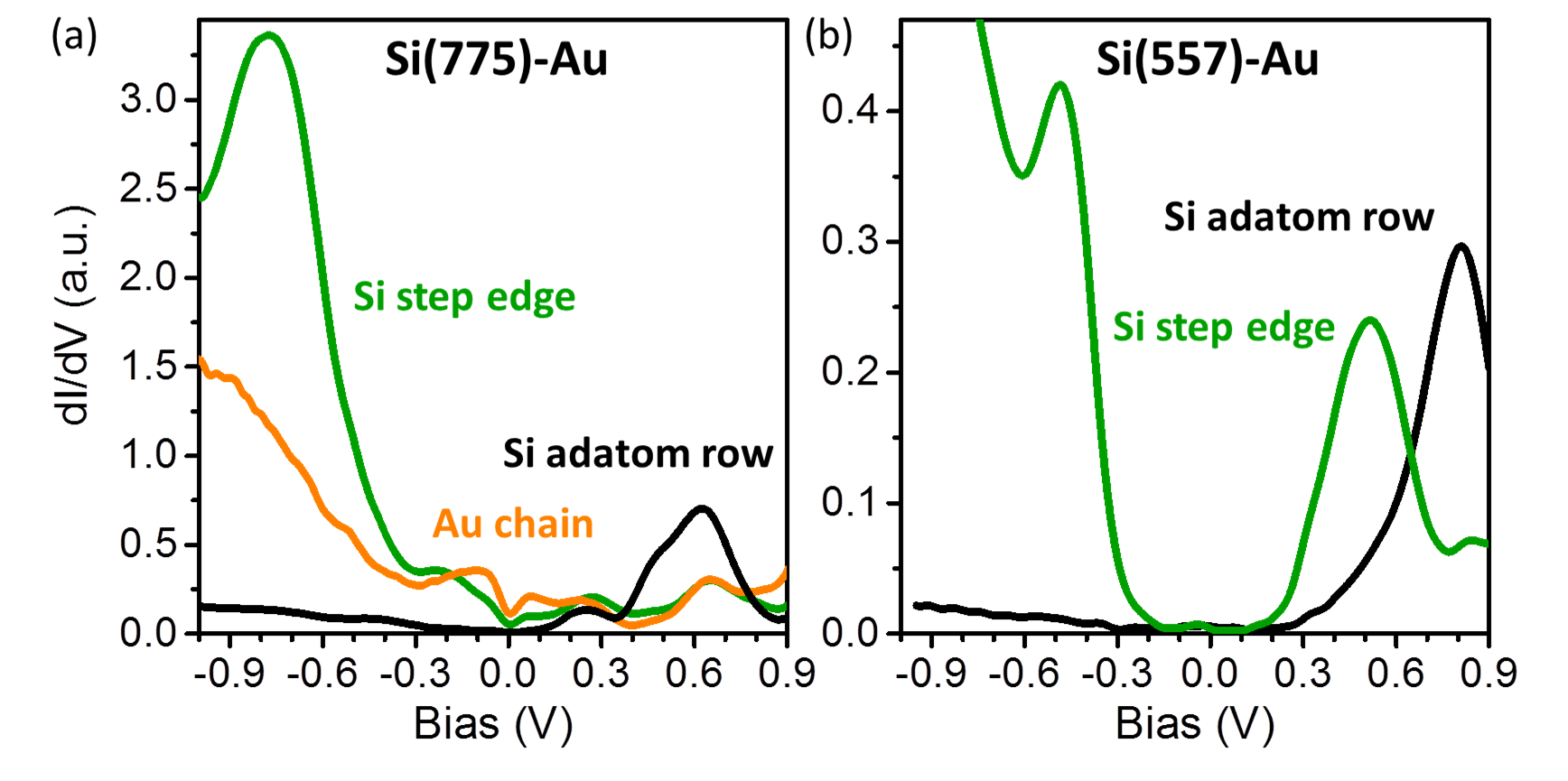}
 \caption{(a) Scanning tunneling spectroscopy (STS) spectra taken at
   the three different structural elements of the Si(775)-Au surface
    at 77 K.  Several single-point spectra were averaged to reduce
   statistical noise.  (b) STS spectrum from the Si(557)-Au surface at
   the same temperature and location as in (a). In contrast to
   Si(775)-Au, there is an intense feature at $0.5$ V above the Fermi
   level---evidence for spin polarization at the step edge.  Note that
   both panels show a peak from the adatom row at 0.6-0.8 eV, which
   originates from the empty dangling-bond orbital. The smaller peaks
   observed in panel (a) from the Si step edge and the Au chain in
   this energy region arise from the finite spatial resolution of the
   STM tip. These spectra are in good qualitative agreement with
   local densities of states predicted by DFT; see Figure\ S2 in the Supporting Information.}
   \label{fig:STS} 
\end{figure}

These findings are strikingly different from those for Si(553)-Au. In
that case, previous DFT calculations predicted the existence of an unoccupied
electronic state localized at the step edge and several tenths of an eV above the
Fermi level \cite{Erwin2010}. This state was subsequently observed in
STS measurements \cite{Snijders2012,Aulbach,YeomACSNano2015} and may
be considered the experimental fingerprint for a spin-polarized step-edge
state. The absence of such a fingerprint for Si(775)-Au calls for an
explanation of this difference. We will turn to this explanation
below. But first we present results for one more family member,
Si(557)-Au.

{\bf Experimental Evidence for Spin Polarization in Si(557)-Au.}  The Si(557)-Au
surface offers an interesting crystallographic contrast to those on
Si(553)-Au and Si(775)-Au, because its steps are oriented oppositely
\cite{Crain2004}. A complete structural model has already been
published and is largely consistent with earlier X-ray data
\cite{Sanchez,Robinson}. This model is again composed entirely of
structural motifs from Figure\ 1a. Nothwithstanding its different
crystallographic orientation, the step edge is again a
single-honeycomb graphitic strip of silicon. Despite the similarity of
its Miller index to (775), the reversed orientation implies a shorter
terrace, $T=18.8$ \AA. Hence the Si(557) terrace can only accommodate
a single Au chain, which does not dimerize. The remainder of the
terrace is identical to that of Si(775)-Au, consisting of a staggered
double row of adatoms and restatoms.

The theoretically predicted ground state of Si(557)-Au, already
reported \cite{Erwin2010}, shows that every second dangling bond on
the step edge is only singly occupied and hence fully spin
polarized. This ordering of the charge does not by itself induce
1$\times$2 periodicity, because the adatom-restatom row already does
this. For this reason, Si(557)-Au does not undergo any structural
transitions at low temperature to higher order periodicity. In contrast, for the
case of Si(553)-Au the experimental observation of such a transition
(to tripled periodicity) constituted the first compelling evidence for
the existence of spin chains in that system
\cite{Ahn2005,Snijders2006,Erwin2010}. Because Si(557)-Au does not
exhibit this transition, the experimental confirmation of spin chains
in Si(557)-Au must rely on a different approach.

\begin{figure*}
\includegraphics[width=\linewidth]{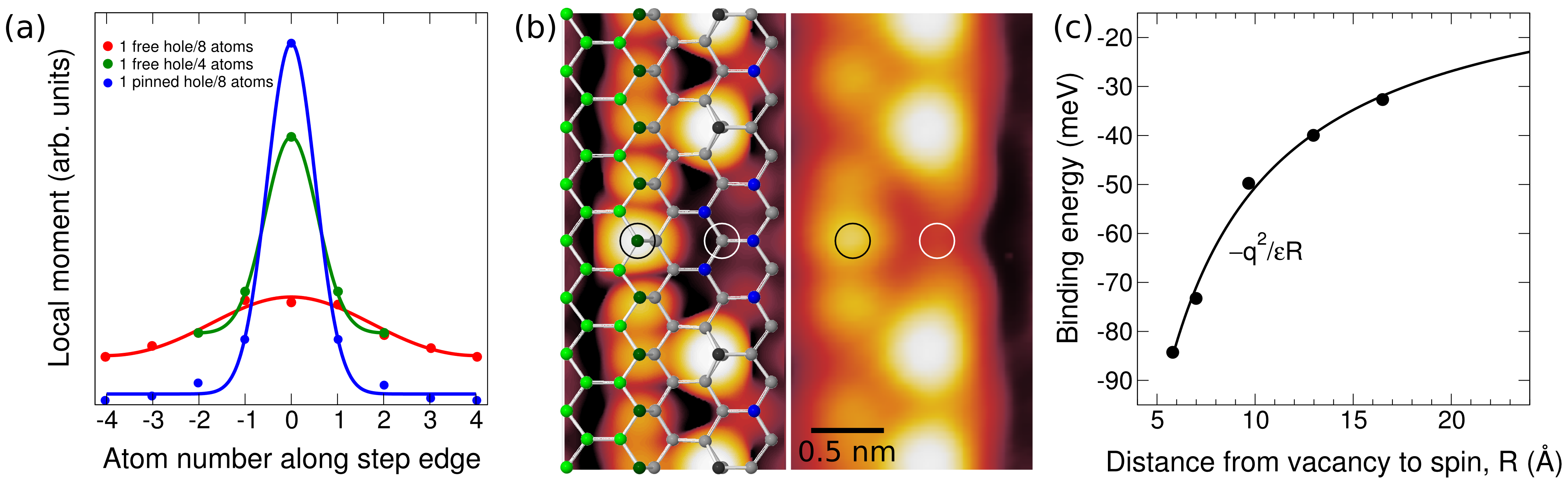}
 \caption{(a) Theoretically predicted formation of spin moments
 when holes are added to Si(775)-Au. When added at low
 concentrations, the
 resulting spins partially delocalize along the step edge; compare the
 cases of one free hole per 8 step-edge atoms (red) and per 4 step-edge atoms (green).
 Holes can also be pinned by their electrostatic attraction to a
 nearby charged defect [here, the adatom vacancy in panel (b)]. In this case the
 spin is more strongly localized (blue). Curves are Gaussian fits to DFT local magnetic moments.
 (b) Simulated and
 experimental empty-state STM images of an adatom vacancy (white
 circle) on Si(775)-Au. The vacancy adds one hole, which renders the
 nearest step-edge atom (black circle)
 spin-polarized and hence brighter in STM. (c) Theoretical
 electrostatic binding energy of the negatively charged adatom vacancy
 to the positively charged step-edge spin. The lowest energy
 configuration, with binding energy 85 meV, is realized in (b).}
 \label{fig:Defect} 
\end{figure*}

Here we present experimental evidence that supports the theoretical
prediction of an unoccupied state, at several tenths of eV above the
Fermi level, arising from spin polarization of a step-edge atom on
Si(557)-Au.  Figure 3b shows the $dI/dV$ spectrum from the step
edge. A well-defined state at 0.5 eV above the Fermi is clearly
evident \footnote{The small spacing and STM-broadened appearance
of the half-filled dangling-bond orbitals make it technically infeasible to
separately resolve the spectra for the two adjacent silicon atoms.}. This is the same unoccupied dangling-bond state that
was previously observed on Si(553)-Au, and supports the predicted spin
polarization of the step edge of Si(557)-Au.

{\bf Spin Chains and Electron Counting.}  We turn now to understanding
in detail the relationship between the structural motifs on
Si($hhk$)-Au surfaces and the spin-polarization state of their
steps. Each atom on the step edge forms three covalent bonds and thus
has a single $sp^3$ orbital that is not part of any bond. To a first
approximation---neglecting band dispersion---this orbital can be
occupied by either zero, one, or two electrons (as a lone pair). DFT
results for all known Si($hhk$)-Au surfaces show one-electron
occupancy always occurs with a large on-site exchange splitting (of
order 1 eV) irrespective of any magnetic order in the system. This
leads to our principal finding: {\it the condition for forming local spin
  moments is that the electron count is large enough to singly occupy
  some, but not large enough to doubly occupy all, of the step-edge
  orbitals.}  On Si(553)-Au, two-thirds of the orbitals are doubly
  occupied and one-third are singly occupied, forming spin chains with
  period 3$a_0$. On Si(557)-Au, one-half are doubly occupied and
  one-half are singly occupied, forming spin chains with period
  2$a_0$. On Si(775)-Au, every orbital is doubly occupied and so spin
  chains do not form.

Another interesting case is provided by a fourth system, Si(335)-Au,
which has been extensively studied by other researchers. STM
topographic measurements show,\cite{krawiec2005a} and DFT
calculations confirm,\cite{krawiec2013a} that half the step-edge
orbitals are doubly occupied and half are empty. DFT calculations
also show that local spin moments do not form on Si(335)-Au, consistent with the
condition proposed above. 
In the Supporting Information we show that for all four Si($hhk$)-Au
surfaces these step-edge occupancies, which were obtained from DFT
calculations, can also be easily derived using simple electron counting.

Of course, one might ask why Si(335)-Au does not instead form a spin
chain with both atoms singly occupied. This scenario might appear
plausible because energy is gained by the exchange splitting of
the singly occupied atoms. But there is a large Coulomb energy penalty for having
{\it adjacent} atoms spin-polarized; this was previously discussed
in a theoretical investigation of finite-temperature dynamics of the
spins on Si($hhk$)-Au. \cite{Erwin2013} Hence this hypothetical
scenario is energetically forbidden.

The completely non-polarized step of Si(775)-Au provides an
interesting test case for the notion of creating and manipulating
spins in Si($hhk$)-Au more generally. Our picture of electron counting
suggests that if one adds holes to Si(775)-Au then local spin moments
would be created there, provided that the holes localize at the
step edge and not somewhere else. Figure 4 demonstrates and
illuminates this result in several ways.  We begin by adding 
holes to the ideal Si(775)-Au system and analyzing their effect. Our
findings are as follows. (1) The holes indeed completely localize at
the step edge. (2) Local magnetic moments develop at step-edge atoms,
with the total moment equal to the number of added holes. (3) These
spin moments are partially delocalized along the step edge, as shown
in Figure\ 4a. (4) The degree of spin delocalization depends on the
hole concentration, unless charged pinning defects are present. In
that case electrostatic attraction between the spin and the defect
further localizes the spin along the step; we will return to this
below. Based on this theoretical proof of principle, we turn now to
explicit chemical strategies for adding holes (or electrons) to
Si($hhk$)-Au surfaces and hence creating spins.

{\bf Spin Chains and Surface Chemistry.} We propose that surface chemistry
offers a potentially useful tool for creating (or suppressing) spin
chains in the family of Si($hhk$)-Au surfaces. The basic idea is to
use native defects and adsorbates on the terrace to control the
electronic configuration on the step edge. Gold atoms have already been used for 
doping Si(553)-Au to study electronic confinement \cite{Songdoping}, so it 
is plausible that other adsorbates will work.

The Si(775)-Au surface has several naturally occurring defects with a
surface concentration of 5 to 10\%. Here we demonstrate
that one of these defects---a missing Si adatom from row B---creates one
spin at the step edge.  To provide an intuitive
understanding of this result we first count electrons. Each adatom
brings four electrons to the surface. Three of these are used to form
covalent bonds from the three surface dangling bonds that surround the
adatom site.  Our analysis shows that the single surface orbital of
the adatom itself is empty. Thus the fourth electron must reside
elsewhere. Our DFT calculations show that this electron goes to the
nearest available orbital on the step edge, creating a doubly occupied
lone pair. With this understanding in hand we now reverse the analysis
and obtain the following fundamental result: creating an adatom
vacancy removes one electron from a doubly occupied step edge
orbital. In DFT a singly occupied step-edge orbital is fully spin
polarized. Hence each adatom vacancy creates one spin at the step
edge.

Experimental evidence indeed supports this analysis. Figure 4b shows
a DFT-simulated STM image of the region around an adatom vacancy. The
nearest step-edge atom is now spin-polarized, which increases its
intensity when imaged at positive bias because the single unoccupied
state is above the Fermi level. The experimental STM image shows
exactly this local enhancement of the nearest step-edge atom; see
Supporting Information Figure\ S3 for a bias-dependent analysis.

Each spin created in this way is positively charged relative to the
background of non-spin-polarized step-edge atoms. It follows by
electroneutrality of the overall system that each adatom vacancy is
negatively charged. Hence we expect an electrostatic attraction
$-q^2/\epsilon R$ between the vacancy and the spin. We tested this
hypothesis theoretically by forcing the spin to localize on several
nearby step-edge atoms (which it does metastably) and then computing
the change in the DFT total energy relative to the case of large separation.
The resulting binding energy curve, shown in Figure\ 4c, confirms the
hypothesis. Our experimental STM results are also consistent with this
attractive interaction: Figure\ 4b shows that the spin indeed
localizes at the step edge atom closest to the adatom vacancy.  We
anticipate that for a group of such artifically created spins, the
lowest energy configuration will depend on the distribution of adatom
vacancies and the shape of the resulting electrostatic landscape.

In addition to adatom vacancies, two other defects are commonly found
on Si(775)-Au. Both are misplaced Si adatoms located one unit cell to
the right of the normal position. The first creates a phase shift of
the adatoms by $a_0$ along the row, while the second does not.
Electron counting shows that these defects are electron-donating and 
charge neutral, respectively (see Supporting Information Figure\ S4). Thus these misplaced-adatom defects
do not create step-edge spins but rather destroy them. DFT
calculations confirm, and STM images are consistent with, this
prediction; see the discussion accompanying Supporting
Information Figure\ S4.

This general model of spin creation and destruction by hole and
electron doping, respectively, motivates our proposal for using
surface chemistry to create and manipulate spin chains.  For example,
one can envision using atom manipulation techniques to create and
arrange adatom vacancies on Si(775)-Au, and hence to construct spin
chains with tunable lengths and spacings. In addition to the
native defects discussed here, a large class of foreign adsorbates
offers broad opportunities---depending on their electronic
character---for inducing or suppressing step-edge spins at
Si($hhk$)-Au surfaces.

Finally, it is likely that our results and proposal for controlling
spins at the steps of Si($hhk$)-Au are also relevant---probably with
interesting modifications---to many other related materials systems,
such as silicene or germanene \cite{Lalmi2010, Vogt2012, Feng2012,
Davila2014}, as well as to other physical
configurations, such as finite ribbons.

\begin{flushleft}{\bf \large Methods}\end{flushleft} {\bf
Experimental.} The $n$-doped (phosphorus) Si(775) and Si(557)
substrates were cleaned by direct current heating up to 1260
$^\circ$C. During Au evaporation (0.32 ML for Si(775)-Au and 0.18 ML
for Si(557)-Au) the sample was held at 650 $^\circ$C. In contrast to
Si(553)-Au \cite{Crain2003} post-annealing was not
necessary. Successful preparation was checked with low-energy electron
diffraction.  High-resolution STM and STS measurements were performed
with an commercial Omicron low-temperature STM at a sample temperature
of 77 K.  Spectroscopy data were obtained via the lock-in detection
method using a modulation voltage of 10 meV (20 meV for Si(557)-Au
data) at a frequency of 789 Hz.  Lock-in $dI/dV$ spectra have been
offset-corrected using simultaneously recorded $I(V)$
curves as a reference.

{\bf Theoretical.} First-principles
total-energy calculations were used to determine the relaxed
equilibrium geometry and electronic properties of the structural model
for Si(775)-Au.  The calculations were performed using a
hydrogen-passivated slab with four silicon double layers plus the
reconstructed top surface layer and a vacuum region of at least 10
\AA.  All atomic positions were relaxed, except the bottom Si layer
and its passivating hydrogen layer, until the largest force component
on every atom was below 0.02 eV/\AA. Total energies and forces were
calculated within the generalized-gradient approximation of Perdew,
Burke, and Ernzerhof \cite{perdew1996} to DFT using
projector-augmented wave potentials as implemented in VASP
\cite{kresse1996}. Results obtained using the local-density
approximation (LDA) are very similar.
In the Supporting Information we also show theoretical
results for the local density of states of Si(775)Au and Si(557)Au,
calculated using the Heyd-Scuseria-Ernzerhof (HSE) hybrid functional.
\cite{heyd_j_chem_phys_2003a,heyd_j_chem_phys_2006a}
The plane-wave cutoff for all calculations was 350
eV.  The sampling of the surface Brillouin zone was chosen according
to the size of the surface unit cell; for the 1$\times$2
reconstruction of Si(775)-Au shown in Figure\ 2a we used 2$\times$4
sampling.  Simulated STM images were created using the method of
Tersoff and Hamann \cite{PhysRevB.31.805}. Spin-polarization, but
not spin-orbit coupling, was included in all the calculations.

\begin{flushleft}{\bf Notes}\end{flushleft} 
\begin{flushleft}
The authors declare no competing financial interest.\\ 
\end{flushleft} 

Supporting Information Available: Detailed discussions of (1)
electron counting of silicon step-edge occupancies on Si($hhk$)-Au;
(2) misplaced Si-adatom defects and their role in electron
transfer. Also included are four additional figures with STM images and theoretical
densities of states.

\begin{flushleft}{\bf Acknowledgements}\end{flushleft} 
The authors thank F.J. Himpsel for many helpful discussions.  This work
was supported by Deutsche Forschungsgemeinschaft (through Grants SFB 1170 ``TocoTronics''
and FOR 1700) and the Office of Naval Research through the Naval
Research Laboratory's Basic Research Program (SCE). Computations were
performed at the DoD Major Shared Resource Centers at AFRL.

\bibliography{775_JA2}

\pagestyle{plain}
\renewcommand{\figurename}{\textbf{Figure}}
\renewcommand{\thefigure}{\textbf{S\arabic{figure}}}
\onecolumngrid
\clearpage
\setcounter{equation}{0}
\setcounter{figure}{0}
\setcounter{table}{0}

\begin{center}
\textbf{{\sffamily {\mdseries Supporting Information}
\\
\vspace{0.5cm}
Spin Chains and Electron Transfer at Stepped Silicon Surfaces}} 
\end{center}

\setcounter{page}{1}


\begin{flushleft}{\bf 1. Electron counting of silicon step-edge occupancies on Si($hhk$)-Au surfaces}\end{flushleft}

\parindent 0in
In the main text we gave the results of DFT calculations for the
silicon step-edge occupancies on four Si($hhk$)-Au surfaces as follows:

\begin{quote}
On Si(553)-Au, two-thirds of the orbitals are doubly
  occupied and one-third are singly occupied, forming spin chains with
  period 3$a_0$. On Si(557)-Au, one-half are doubly occupied and
  one-half are singly occupied, forming spin chains with period
  2$a_0$. On Si(775)-Au, every orbital is doubly occupied and so spin
  chains do not form.  On Si(335)-Au, one-half the step-edge
  orbitals are doubly occupied and one-half are empty and so spin
  chains do not form. These results are summarized in this table:
\begin{center}
\begin{tabular}{l|c|c}
  \hline
  Surface & Individual occupancies & Average occupancy\\
  \hline
  Si(335)-Au & 2, 0    & 1     \\
  Si(557)-Au & 2, 1    & 3/2   \\
  Si(553)-Au & 2, 2, 1  & 5/3   \\
  Si(775)-Au & 2      & 2     \\
  \hline
\end{tabular}
\end{center}
\end{quote}

\parindent 0in
Here we show how the average occupancies can be easily obtained using
simple electron counting. From these average occupancies it is 
straightforward to deduce the individual occupancies---and hence both the existence
and period of spin chains on arbitrary Si($hhk$)-Au surfaces.

\parindent .5in
In the {\it Background and Preliminaries} section of the main text, we
showed that the hypothetical ``ideal'' step
edge of Si($hhk$)-Au surfaces---that is, before any electron transfer
to or from the terrace occurs---consists of doubly occupied lone pairs
on every silicon step-edge atom. This result was obtained by
considering the electronic structure of Si(111)-$M$ surfaces (where $M$
is an alkali), for
which there is no step. When a step and terrace are introduced the electron
counting is slightly more complicated. In the following we apply
this counting to the four surfaces in the table.

We begin with the simplest system, Si(335)-Au, which has the smallest
possible terrace, consisting of a single silicon row. Each atom in
this row has two dangling bonds which combine to form two bands, one
bonding and one antibonding. The bonding band is completely filled and
the antibonding band (which has parabolic dispersion and is
traditionally called the ``Au band'') is approximately half-filled in
DFT calculations; hence a total of three electrons is required. Two
come from the singly occupied dangling bonds themselves and one from
the Au atom. Because the Au electron is no longer available to occupy
the step-edge orbital, this reduces the occupancy of the step-edge
orbital from two to one. Hence the average occupancy of a Si(335)-Au
step-edge orbital is 1.

The Si(557)-Au surface is slightly more complicated because it
contains, in addition to the features of Si(335)-Au, one row of
silicon adatoms with 1$\times$2 periodicity. In the main text it was
shown that each such adatom adds one electron to the system. Hence,
because there is one adatom for every two step-edge atoms, the
average occupancy of a Si(557)-Au step-edge orbital is 1/2 more than
for Si(335)-Au, that is, 3/2.

The last two surfaces, Si(553)-Au and Si(775)-Au, are still more
complicated because the Au chain is now double rather than single.
Si(553)-Au has no adatoms and, aside from the double Au chain, is
structurally equivalent to Si(335)-Au. Thus we begin the electron
counting from that reference point. Each Au atom in the second chain
breaks an intact Si bond on the terrace of the Si(335)-Au surface,
leaving a row of Si dangling bonds which form another parabolic ``Au
band'' which is one-sixth filled; hence one-third of an electron is
required. The Au electron is needed to passivate another surface Si
dangling bond created at the bottom of the step edge, and thus the
required one-third electron must come from the electron in the
terrace-Si dangling bond. This leaves two-thirds of an electron left
over to occupy the step-edge orbital. Hence the average occupancy of a
Si(553)-Au step-edge orbital is 2/3 more than for Si(335)-Au, that is,
5/3.

We count electrons on Si(775)-Au starting from the reference
state Si(557)-Au because both surfaces are, aside from the Au chain, 
structurally equivalent. Each Au atom in the second chain
breaks an intact Si bond on the terrace of the Si(557)-Au surface,
which leaves a row of Si dangling bonds to form another parabolic ``Au
band'' which is one-fourth filled; hence one-half of an electron is
required. As argued above, this must come from the electron in the
terrace-Si dangling bond. This leaves one-half electron 
left over to occupy the step-edge orbital. Hence the average occupancy of a
Si(775)-Au step-edge orbital is 1/2 more than for Si(557)-Au, that is,
2.

Finally, from these average occupancies we can deduce the actual
individual occupancies. Two rules suffice, both explained in the main
text: (1) Only integer occupancies are allowed; (2) Singly
occupied orbitals cannot be adjacent. Applying these rules we find
that Si(335)-Au has 1$\times$2 periodicity with individual occupancies
(2, 0); Si(557)-Au has 1$\times$2 periodicity with occupancies (2, 1);
Si(553)-Au has 1$\times$3 periodicity with occupancies (2, 2, 1); and
Si(775)-Au has 1$\times$2 periodicity (due to the adatom row) with
occupancies (2,2). These periodicities and occupancies agree with
those obtained from detailed DFT calculations and are consistent with
all available experimental data. Moreover, the onsite exchange
splitting of singly occupied orbitals implies the existence of spin
chains for Si(553)-Au and Si(557)-Au, but not for Si(335)-Au or
Si(775)-Au; this is also consistent with DFT calculations and
experiment.

\begin{flushleft}{\bf 2. Misplaced Si-adatom defects and electron transfer}\end{flushleft}

In addition to the adatom vacancy discussed in the main text (Figure 4b),
the Si(775)-Au surface hosts two other common defects namely two
different types of misplaced adatoms. A structural model and simulated
STM images, as well as experimental STM data for both misplaced adatom
defects (Type I and Type II), are displayed in
Figures S4a and b, respectively. In both cases the
theoretical simulation is in detailed agreement with the experimental
STM images, giving excellent support for both structural
models. Defect Type I interrupts the usual 2$a_{0}$ spacing of normal
adatoms, reducing by $a_{0}$ the normal 4$a_{0}$ spacing between the
two adjacent normal adatoms. As indicated by the colored dots and
stars in the unoccupied STM images at 0.5 and 0.2 V not only the
adatom row, but also the Au chain and the Si step edge atoms,
exhibit this phase shift by $a_{0}$ (see Figure S4a).

\parindent .5in
The effect of this native defect on the step edge's electron
concentration is easiest to evaluate in three stages: As discussed in
the main text, removal of a normal adatom and its four electrons from
the model in Figure 2a transfers one hole to the step edge. Next we
shift the entire row of normal adatoms on one side of this vacancy
toward it by $a_{0}$, anticipating the actual spacing around the
misplaced adatom defect. This increases by one-half the total number
of adatoms. Since the addition of half an adatom is equivalent to the
addition of half an electron, we have so far added 0.5 holes to the step
edge. Finally, we return the adatom to its new misplaced site,
passivating two silicon atoms and forming one bond to a gold atom;
thus this returned adatom is again equivalent to adding one
electron. Considering all three stages, we see that the overall net
change created by a misplaced adatom defect is \textit{0.5 additional
electrons}. Thus, the misplaced adatom defect Type II should \textit{not}
induce any spin-polarization in the step edge atoms. This is confirmed
by the experimental STM images around adatom defect Type I, which do
not show any evidence of spin-polarized step-edge atoms. 

\parindent .5in
For misplaced adatom defect Type II (Figure S4b) the
electron counting follows the same line as for Type I. However, the
1$\times$2 cell lacks the additional adatom (and thus half an
electron) induced by the phase shift of $a_{0}$, which occurs only for defect
Type I. Therefore, defect Type II is \textit{charge neutral} and
should not induce any spin polarization on step-edge atoms. Again, this is
consistent with the experimental STM results and hence confirms our
description of electron transfer between native defects and the
silicon step edge.

\newpage
\begin{flushleft}{\bf 3. Supplementary figures}\end{flushleft}

\begin{figure}[htbp]
\includegraphics[width=0.85\linewidth]{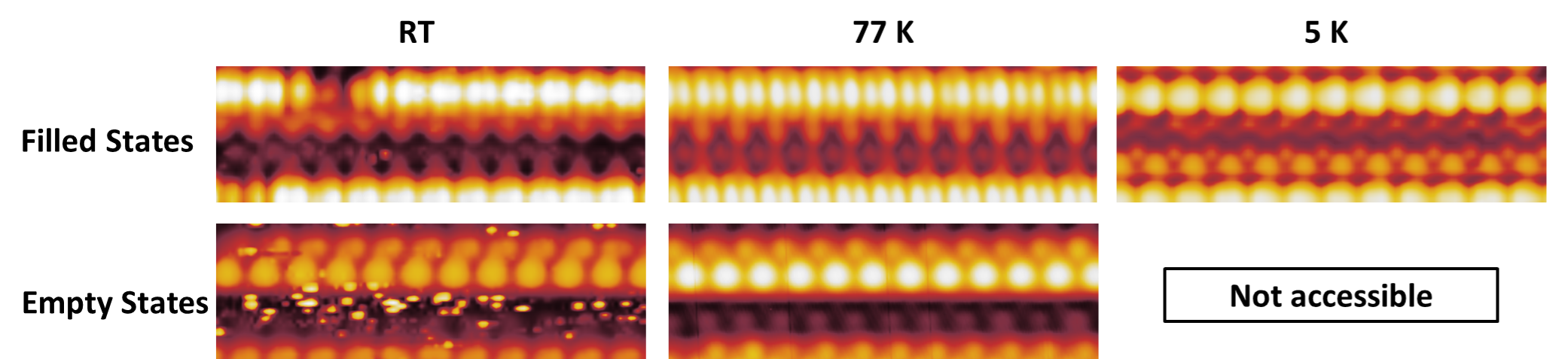}
\caption{\textbf{Absence of a phase transition in Si(775)-Au.} 
Constant current STM images of the Si(775)-Au surface at various sample temperatures. Filled states images have been taken at U~=$-$1.0~V (U~=$-$0.95~V for 77~K) and empty states are shown for U=$+$1.0~V. All three chain types, namely, the Si honeycomb row, the Si adatom row as well as the Au chain, do not change their $\times$2 periodicity between room temperature and 5~K, which indicates the absence of a phase transition in this temperature window.} 
\end{figure}

\begin{figure}[htbp] \includegraphics[width=0.8\linewidth]{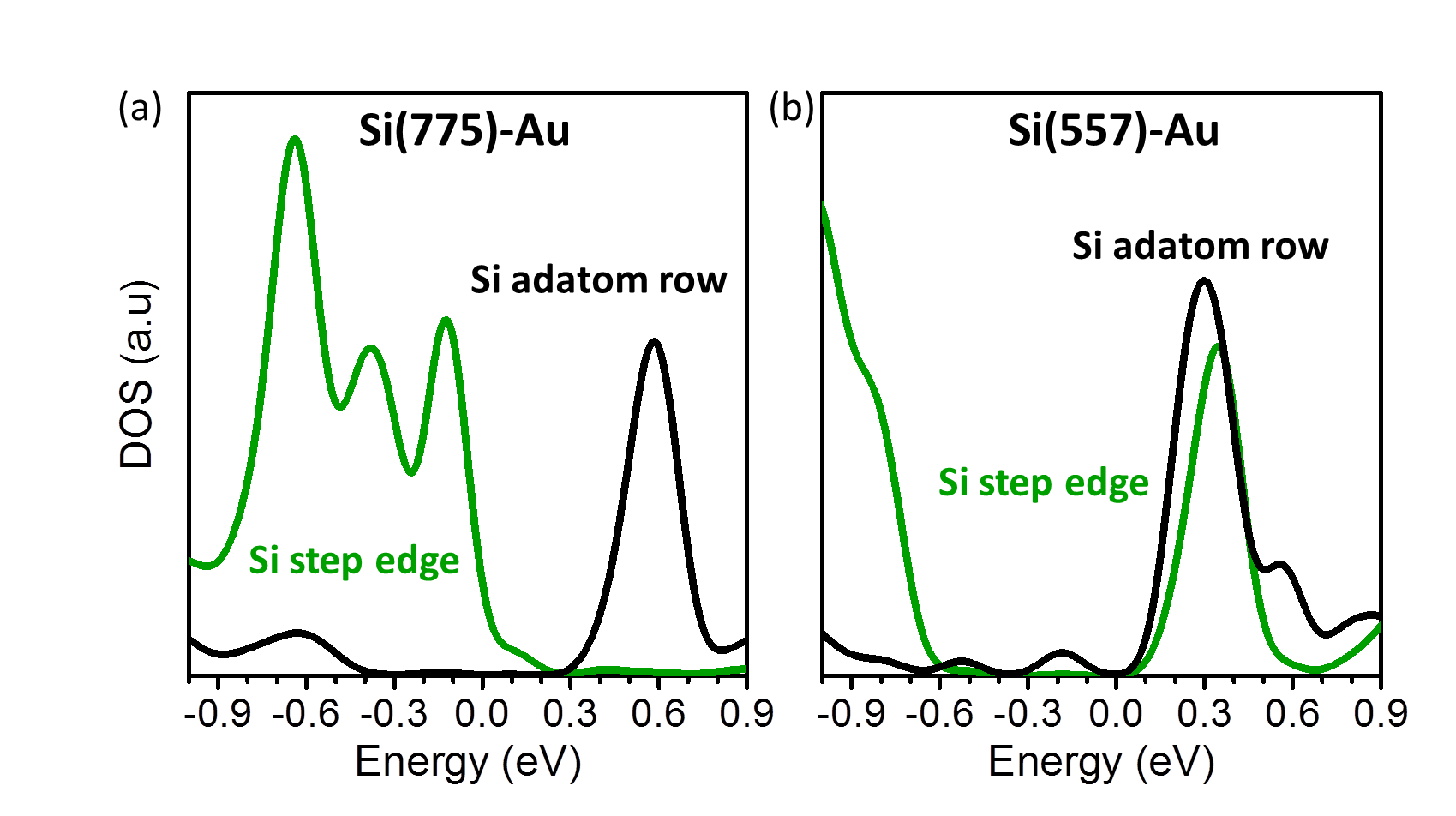}
\caption{\textbf{Theoretical DFT/HSE local densities of states for
Si(775)-Au and Si(557)-Au.} (a) For Si(775)-Au the step edge is
predicted to consist of doubly-occupied lone-pairs on every
atom. Hence the step-edge local density of states (LDOS) shows no
states above the Fermi level; only the unoccupied orbitals of the Si
adatom row appear there. (b) For Si(557)-Au the step edge is predicted
to consist of alternating doubly-occupied and singly-occupied
atoms. Hence the LDOS additionally shows a strong peak above the Fermi
level. Both panels are in good qualitative agreement with Figure 3 of
the main text.} \label{fig:defects} \end{figure}

\begin{figure}[htbp] \includegraphics[width=0.85\linewidth]{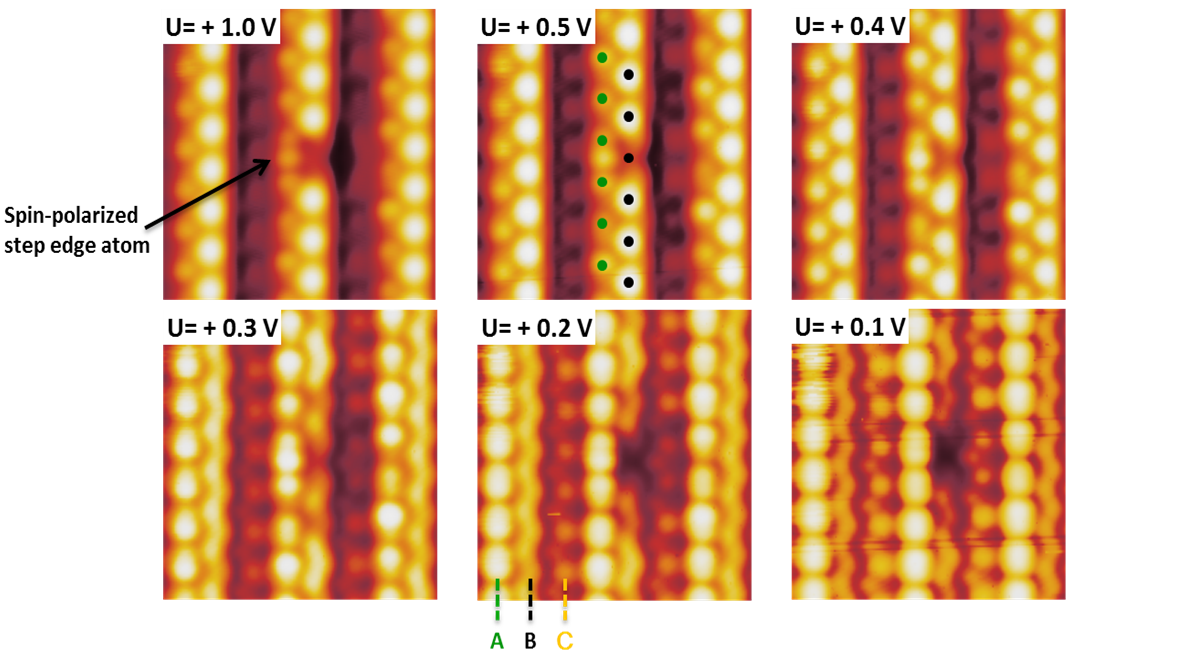}
\caption{\textbf{Bias dependence of an adatom vacancy.}  STM bias
series of the Si(775)-Au unoccupied states including an adatom
vacancy. The adatom vacancy dopes one hole to the step edge and thus
renders the nearest step-edge atom (black arrow) spin-polarized and
hence brighter in STM. The green and black dots make it clear that the
this vacancy does not create a phase shift of other features within
the same row (in contrast to the misplaced adatom defect in
Figure S4a). These images are consistent with the structural model of a
vacancy shown in Figure 4b of the paper. In addition, the zigzag
appearance of row B at bias $\leq$ 0.3 V, as well as the inversion of
adatom vs honeycomb row relative intensities that occurs at about
0.4~V, provide additional support for our basic structural model of
Si(775)-Au.} \end{figure}
 
\begin{figure}[htbp]
\includegraphics[width=0.8\linewidth]{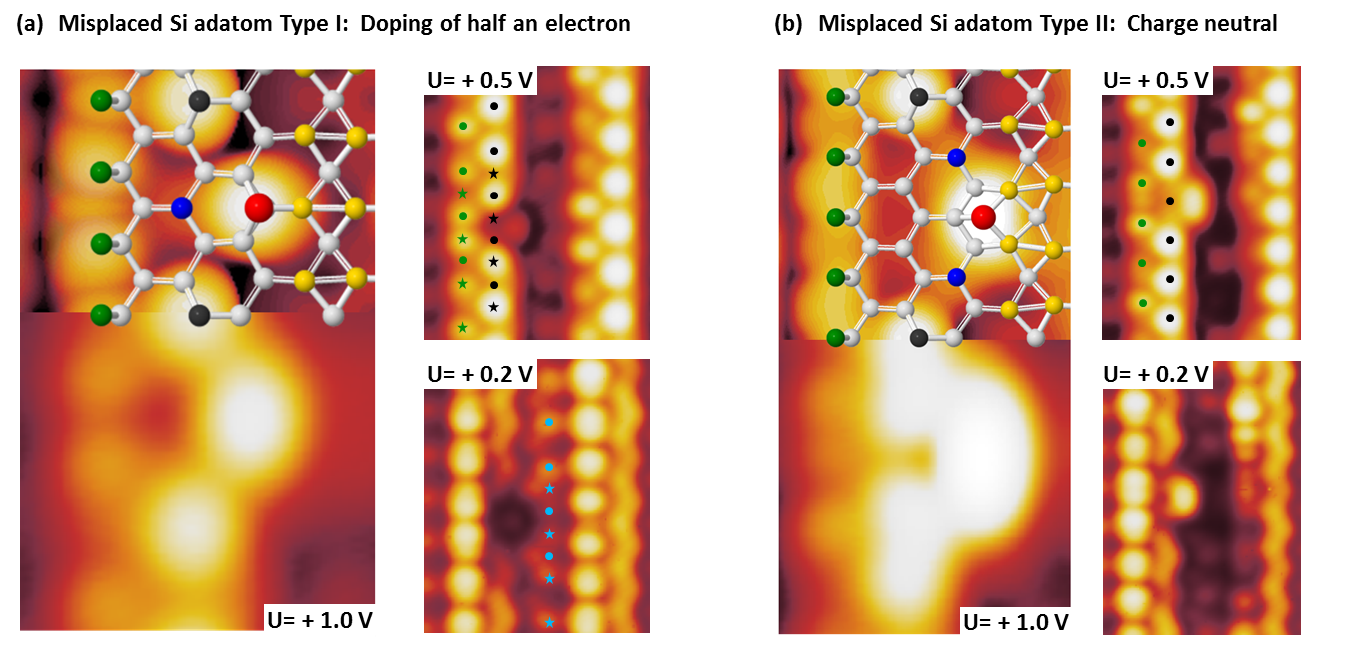}
\caption{\textbf{Two types of misplaced-adatom defects and the resulting charge transferred to the silicon step edge.}
(a) Structural model and STM imagery for Type I misplaced-adatom defect. This defect disrupts the usual adatom spacing, as indicated by the colored dots and stars in the right-hand panels. (b) Structural model and STM imagery for Type II misplaced-adatom defect. In contrast to the Type I defect, the Type II defect does not disrupt the adatom spacing, as demonstrated by the green and black dots in the right-hand panel.} \label{fig:defects}
\end{figure}

\end{document}